\documentclass[         %
aps,                    %  American Physical Society
prc,                    %  Physical Review
showpacs,               %  Displays PACS after abstract
superscriptaddress,     %  Authors' addresses linked with superscripts
nofootinbib,            %  Does not treat footnotes as references
twocolumn,              %
]
{revtex4}               %  REVTEX 4 Package used
\usepackage{graphicx}
\usepackage{amsmath}
\usepackage{epstopdf}

\begin{document}

%%%%%%%%%%%%%%%%%%%%%%%%%%%%%%%%%%%%%%%%%%
%%%%%%%%%%%%%%%%%%%%%%%%%%%%%%%%%%%%%%%%%%

\title{Direct Measurement of Supernova Neutrino Emission Parameters\\
with a Gadolinium-Enhanced Super-Kamiokande Detector}

\author{Hasan Y{\"u}ksel}
%\email{yuksel@mps.ohio-state.edu}
\affiliation{Department of Physics, Ohio State University, Columbus,
Ohio 43210, USA}

\author{Shin'ichiro Ando}
%\email{ando@utap.phys.s.u-tokyo.ac.jp}
\affiliation{Department of Physics, School of Science, University of
Tokyo, Tokyo 113-0033, Japan}

\author{John F. Beacom}
%\email{beacom@mps.ohio-state.edu}
\affiliation{Department of Physics, Ohio State University, Columbus,
Ohio 43210, USA}
\affiliation{Department of Astronomy, Ohio State University, Columbus,
Ohio 43210, USA}

\date{12 September 2005; minor revisions 26 May 2006}

\begin{abstract}
The time-integrated luminosity and average energy of the neutrino emission
spectrum are essential diagnostics of core-collapse supernovae.  The
SN 1987A electron antineutrino observations by the Kamiokande-II and
IMB detectors are only roughly consistent with each other and theory.
Using new measurements of the star formation rate history, we
reinterpret the Super-Kamiokande upper bound on the electron
antineutrino flux from all past supernovae as an excluded region in
neutrino emission parameter space.  A gadolinium-enhanced
Super-Kamiokande should be able to jointly measure these parameters,
and a future megaton-scale detector would enable precision studies.
\end{abstract}

\pacs{97.60.Bw, 98.70.Vc, 95.85.Ry, 14.60.Pq
\vspace*{-0.3cm}}

% 97.60.Bw Supernovae
% 98.70.Vc Background radiations
% 95.85.Ry Neutrino, muon, pion, and other elementary particles; cosmic rays
% 14.60.Pq Neutrino mass and mixing

\maketitle

%%%%%%%%%%%%%%%%%%%%%%%%%%%%%%%%%%%%%%%%%%
%%%%%%%%%%%%%%%%%%%%%%%%%%%%%%%%%%%%%%%%%%

%{\bf Introduction.---}%
%
When a massive star dies, its core collapses and rebounds, producing
an outgoing shock wave that should eject the stellar envelope,
causing the optical supernova, and leaving behind a neutron star
remnant.  However, in simulations, the shock wave stalls, leading to
the whole star collapsing into a black hole, failing to produce an
optical supernova or spread its heavy-element yields~\cite{SNsim}.
Since the required explosion energy is only $\sim 1\%$ of the emergent
neutrino energy, a full accounting of the neutrino emission is
essential for understanding supernovae.  Further, in the Bethe-Wilson
delayed explosion model, the neutrinos revive the
shock~\cite{delayed}.  Resolution of the supernova problem would also
have profound implications for the history of stellar evolution and
nucleosynthesis.

The weak interactions of neutrinos, which allow them to reveal the
dynamics deep within the exploding star, also make their detection
challenging.  The last nearby supernova, SN 1987A, occurred in the
Large Magellanic Cloud at 50 kpc, and $\simeq 20$ neutrinos were
detected~\cite{1987ADETECT} preceding the optical supernova,
confirming our basic understanding of the
explosion~\cite{1987Areviews}.  However, even taking into account the
small statistics, the fitted ranges for the time-integrated luminosity and
average energy are perplexing, showing clear discrepancies among the
experimental detections and theory~\cite{1987Afit1,1987Afit2}.  A
Milky Way supernova would yield many events in present detectors, but
the expected supernova rate is only $\sim 3$ per century.  We have
shown that with proposed megaton-scale detectors, it will be possible
to build up the spectrum by detecting neutrinos one or two at a time
from supernovae within 10 Mpc, at a rate as large as $\sim 1$ neutrino
per year~\cite{PEEPING}.

%---------------------------------------------------------------------------%
\begin{figure}
\includegraphics[width=3.25in,clip=true]{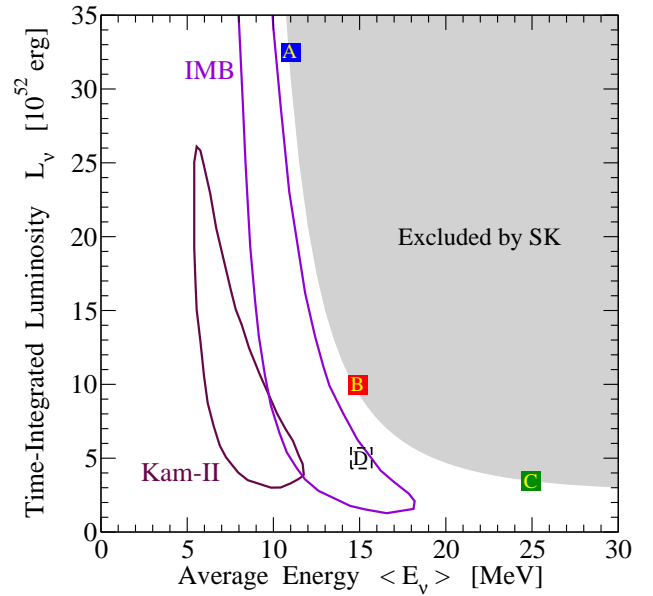}
\caption{(Color online)
Joint limits on the time-integrated luminosity $L_{\nu}$ and the
 spectrum average energy $\langle E_\nu \rangle$ for electron
 antineutrinos.  The two contours are the allowed regions at 90\%
 C.L. from the SN 1987A analysis of Ref.~\cite{1987Afit1} and our
 shaded region corresponds to the SK 90\% C.L. upper limit on the DSNB
 flux~\cite{Malek}.
\label{fig:temp-limit}}
\vspace{-0.25cm}
\end{figure}
%---------------------------------------------------------------------------%

Here we propose a new approach, which could begin immediately, if the
existing Super-Kamiokande (SK) detector were modified by the addition
of gadolinium to greatly reduce backgrounds, as proposed by Beacom and
Vagins~\cite{GADZOOKS,Vagins}.  We consider the spectrum of the Diffuse
Supernova Neutrino Background
(DSNB)~\cite{AST,DSNB,Ando2004,Concordance} as the observable.  The
DSNB predictions depend on the redshift evolution of the supernova
rate, which is separately measurable and increasingly well known, and
the neutrino emission per supernova, the object of our study.  While
the received neutrino spectrum will be redshifted, it will have
relatively high statistics, up to several events per year in SK.
Recently, the DSNB uncertainties from the star formation rate
history~\cite{GALEX,Hopkins} narrowed enough that it is now sensible to
reinterpret the SK flux limit~\cite{Malek} as an exclusion
region in the plane of the time-integrated luminosity and average energy,
which can be directly compared to the allowed regions from SN 1987A. 
Anticipating further improvements in the astronomical data, we show that a
gadolinium-enhanced SK should be able to usefully constrain the emission
parameters in much of the interesting range.  

%%%%%%%%%%%%%%%%%%%%%%%%%%%%%%%%%%%%%%%%%%

{\bf Supernova 1987A Signal.---}%
One of the triumphs of astrophysics, nuclear physics and particle
physics was the detection of neutrinos from SN 1987A, so far the only
astrophysical source besides the Sun seen with neutrinos.  The $\simeq
20$ events in the Kamiokande-II (Kam-II) and IMB detectors are assumed
to be mostly inverse beta events, $\bar{\nu}_e + p \rightarrow e^+ +
n$; the cross section $\sigma \sim E_\nu^2$, with the positron carrying
nearly the full neutrino energy~\cite{invbeta}.  The number of detected events 
$N_{det} \sim (L_\nu / \langle E_\nu \rangle) \cdot \langle E_\nu \rangle^2
\sim L_\nu \cdot \langle E_\nu \rangle$, where $L_\nu$ is the time-integrated
luminosity of the electron antineutrinos and $\langle E_\nu \rangle$ is the
average energy of the neutrino emission spectrum.  The average detected
energy $\langle E_{det} \rangle \sim \langle E_\nu \rangle$.  The combination
of these two constraints explains the banana-shaped allowed regions shown
in Fig.~\ref{fig:temp-limit}, taken from the full spectrum analysis of
Ref.~\cite{1987Afit1} (those authors assumed a Maxwell-Boltzmann thermal
emission spectrum).  We show only the 90\% C.L., which is appropriate for
this level of precision, to avoid cluttering the figures.

At least three puzzling features of the SN 1987A data still stand out.
First, the fits to the Kam-II and IMB data for the neutrino emission
parameters barely overlap, due to the disagreement on the
spectra~\cite{1987Afit1,1987Afit2}.  Second, the results disagree with
the canonical expectations~\cite{1987Afit1,1987Afit2}, conservatively indicated
by point \textbf{D} in Fig.~\ref{fig:temp-limit}.  This corresponds to a neutron star
binding energy of $3 \times 10^{53}$ erg, assumed shared equally among the
six flavors, and an effective received $\bar{\nu}_e$ temperature of about 5 MeV.
Both the Kam-II and IMB data allow very high luminosities, perhaps
reflecting a larger neutron star binding energy and/or a violation of its assumed
equipartition among flavors.  Both, but especially Kam-II, allow very low
average energies, especially if neutrino mixing with higher-temperature
flavors is taken into account (i.e., with temperatures possibly as large as
8 MeV).   Third, both the Kam-II and IMB results are in
significant disagreement with model-independent tests of the angular
distributions of the detected events~\cite{invbeta,Costantini,DSNBnue1}.

We emphasize that the Kam-II and IMB results are {\it roughly} consistent
with each other and theory; still, there are puzzling issues raised which cannot
be answered without new data.  Also, all supernovae may not be alike,
and the DSNB results will reveal the {\it average} neutrino emission
parameters of supernovae, possibly more relevant for cosmological
applications.  We therefore stress that the adoption of SN 1987A as a
template for DSNB studies is undesirable.

%%%%%%%%%%%%%%%%%%%%%%%%%%%%%%%%%%%%%%%%%%

%---------------------------------------------------------------------------%
\begin{figure}
\includegraphics[width=3.25in,clip=true]{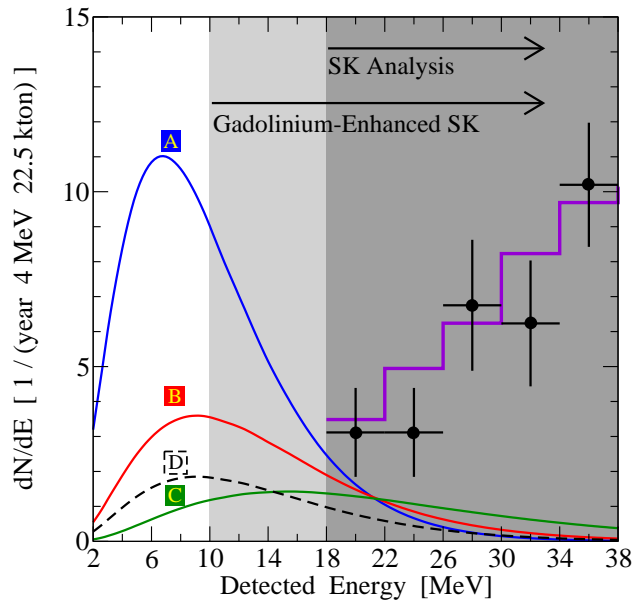}
\caption{(Color online)
DSNB detection spectra for selected
 parameters (solid and dashed curves), efficiency-corrected SK data
 (points with error bars), and detector background (solid steps), all in
 counts per 4 MeV, per year, per SK fiducial volume.  The
 values for the representative points \textbf{A}, \textbf{B},
 \textbf{C} and \textbf{D} are given in Table~\ref{table:temp-table}.
 SK is so far sensitive only to the dark shaded region above 18 MeV
 due to high backgrounds at lower energies (not shown).  With the
 addition of gadolinium, these backgrounds in the range 10--18 MeV
 would be removed, and that shown reduced by a factor $\sim 5$,
 opening up also the light shaded region for analysis~\cite{GADZOOKS,Vagins}.
\label{fig:temp-spect}}
\end{figure}
%---------------------------------------------------------------------------%

{\bf SK DSNB Limit.---}%
To predict the DSNB flux, one needs only the core-collapse supernova
rate as a function of redshift $z$, convolved with the neutrino
emission per supernova, taking into account redshift
effects~\cite{AST,DSNB,Ando2004,Concordance}.  At present, the former
is calculated from the measured star formation rate and the stellar
initial mass function, which determines the fraction of stars that end their
lives as core-collapse supernovae.  We base
our results primarily on the GALEX star formation rate data~\cite{GALEX},
for which the normalization uncertainty is now at the $\simeq 30\%$
level.  This yields results similar to those of the Concordance Model
of Ref.~\cite{Concordance}, which was shown to be consistent with the
latest measured star formation, thermonuclear (type Ia) supernova, and
core-collapse supernova (types II, Ib, and Ic) rates, as well as other
data, and which predicts a DSNB flux just below the present SK limit.
(See also the more recent Ref.~\cite{Hopkins}.)
The DSNB prediction thus depends first on a purely astronomical
factor, which is already measured well and will ultimately be
precisely and unambiguously measured through direct data on the
supernova rate versus redshift (note that optically failed supernovae
with substantial neutrino emission would increase the
core-collapse rate)~\cite{Concordance}.  The second factor, the
average neutrino emission per supernova, must be measured directly,
and this is our focus.

In 2002, SK reported their DSNB flux upper limit for electron antineutrinos
with a detection energy threshold of 18 MeV as 1.2 cm$^{-2}$ s$^{-1}$
(90\% C.L) through non-detection of excess counts above background
fluctuations~\cite{Malek}. Due to the rising background and falling
signal with increasing energy, almost all of the statistical power
derives from the first two bins in Fig.~\ref{fig:temp-spect}, which is
why the SK flux limit was the same for DSNB models with different
spectral shapes.  With the present statistics, it is enough to use
just these two bins to limit the signal, noting that the other bins
fix the background normalization.  While at the time of the SK
analysis, the DSNB models differed significantly in their
normalization, the latest astronomical data greatly restricts this
freedom, and will eventually eliminate it, modulo the differences in
neutrino emission per supernova that we want to test.  Thus it now
makes sense to reinterpret the SK event rate limit ($\simeq 2$
yr$^{-1}$ in 18--26 MeV \cite{Malek}) in terms of the supernova
electron antineutrino emission parameters, the time-integrated luminosity
and average energy.  The result of our analysis is that the shaded
region in Fig.~\ref{fig:temp-limit} is excluded.   (We assumed a thermal
emission spectrum of the form used in Ref.~\cite{1987Afit2}, taking
$\alpha = 3$, which corresponds to a somewhat ``pinched" spectrum.)  This
does not yet reach the allowed regions deduced from the SN 1987A data,
but it is encouragingly close.  Neutrino mixing can blend the initial
$\bar{\nu}_e$ spectrum with higher-energy
$\bar{\nu}_\mu/\bar{\nu}_\tau$ spectra.  We are limiting an effective
composite spectrum, which will be dominated by the harder
spectrum~\cite{AST}.  Thus our analysis is conservative, in that with
neutrino mixing, the interpretation of the DSNB bound would be more
constraining (e.g., Ref.~\cite{Concordance}).  

Beyond the two supernova neutrino emission parameters used here, there
is also the question of the spectrum shape, and whether it is distorted from
thermal by being ``pinched" or ``anti-pinched"~\cite{1987Afit2}.  The SK energy
threshold of 18 MeV is used in Ref.~\cite{Malek} is high, especially noting that
redshifts $z \lesssim 1$ are relevant; the SK limit is thus based on energies
$\simeq 20-40$ MeV in emission, where the spectrum uncertainties are largest.
Indeed, this is part of the motivation for lowering the SK energy threshold,
so that the detected events would correspond to emission from the
better-understood spectrum peak region.

%---------------------------------------------------------------------------%
\begin{table}[t]
\caption{The values for the points \textbf{A}, \textbf{B}, \textbf{C}
 (near the SK upper bound) and \textbf{D} (canonical values) of the
 figures.
\label{table:temp-table} }
\begin{ruledtabular}
\begin{tabular}{lccc}
Point &$\langle E_\nu \rangle$ [MeV] & 
$L_{\nu}$ [$10^{52}$ erg] &  Sensitivity\\
\hline
\textbf{A} &$11$ & $32$ & Average Energy  \\ 
\textbf{B} &$15$ & $10$ & Both Variables \\%
\textbf{C} &$25$ & $3.5$ & Integrated Luminosity \\
\hline
\textbf{D} &$15$ & $5$ & Lowered Sensitivity \\ 
\end{tabular}
\end{ruledtabular}
\end{table}
%---------------------------------------------------------------------------%

One can get more insight by examining three points, \textbf{A},
\textbf{B}, and \textbf{C}, shown in Fig.~\ref{fig:temp-limit} and
Table~\ref{table:temp-table}, which are at the edge of detectability.
We also consider a point \textbf{D}, which is often regarded as the
canonical values for $\bar{\nu}_e$ emission before neutrino mixing.
The DSNB spectra for these points are shown in
Fig.~\ref{fig:temp-spect}, together with the SK data and background
expectations (dominantly from the decays of sub-\v{C}erenkov muons
produced by atmospheric neutrinos).  The three points \textbf{A},
\textbf{B}, and \textbf{C} correspond to almost equal yields
(comparable to the fluctuations in the backgrounds) in 18--26 MeV,
where the present SK sensitivity is greatest.  The point \textbf{D}
produces fewer signal events and is safely allowed.  Note that these
spectra are quite different at lower energies.  Thus it is clear that
in order to make the necessary progress over the present
background-limited search, SK must reduce both the background rates
and the energy threshold.

The points \textbf{A}, \textbf{B}, and \textbf{C}, besides representing
three very different possibilities, are also the most favorable for
being probed by SK in the near term.  (Below, we discuss the outlook
for models with a lower neutrino emission per supernova.)  These
points are chosen just with respect to what is allowed by the DSNB,
deliberately not taking into account other possible constraints, so
that the impact of our results can be clearly seen.  Additionally, it is
important to keep an open mind about what the true parameters are,
given that (a) numerical supernova models fail to explode~\cite{SNsim},
and (b) it remains possible that SN 1987A was very different from an
average supernova.

Points \textbf{A} and \textbf{C} are relatively extreme, compared to the
more canonical point \textbf{D}.  However, possibilities like \textbf{A}
might occur if the neutron star binding energy and the fraction of this
energy carried away by $\bar{\nu}_e$ (i.e., a violation of the usually
assumed equipartition of the energy among the six flavors)
are both larger than expected~\cite{SNsim};  more directly, \textbf{A} is
very close to part of the IMB allowed region for the SN 1987A data. 
Possibilities like \textbf{C} correspond to a temperature of about 8 MeV,
which is within the range considered in many papers, especially if
neutrino mixing is taken into account.
 
%%%%%%%%%%%%%%%%%%%%%%%%%%%%%%%%%%%%%%%%%%

%---------------------------------------------------------------------------%
\begin{figure}[t]
\includegraphics[width=3.25in,clip=true]{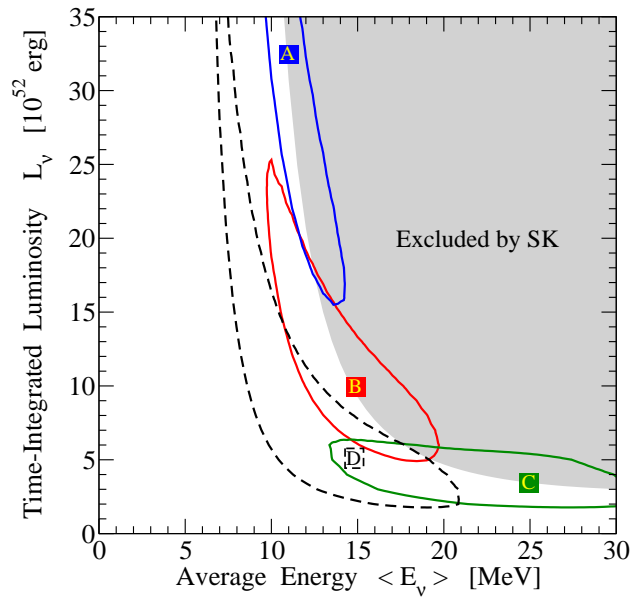}
\caption{(Color online)
Possible 90\% C.L. measurements of the emission parameters of
 supernova electron antineutrino emission after 5 years running of a
 gadolinium-enhanced SK detector.
\label{fig:temp-expect}}
\end{figure}
%---------------------------------------------------------------------------%

{\bf Gadolinium-Enhanced SK Sensitivity.---}%
In SK, DSNB $\bar{\nu}_e$ would be detected by the inverse beta
reaction $\bar{\nu}_e + p \rightarrow e^+ + n$ on free
protons~\cite{invbeta}.  At present, this is a (positron) singles
search, for which there are very large background rates~\cite{Malek}.
With dissolved gadolinium, SK could also detect the neutron via its
radiative capture ($\Sigma E_\gamma \simeq 8$ MeV), which would give a
tight temporal and spatial coincidence for the signal
events~\cite{GADZOOKS,Vagins}.  This would reduce the sub-\v{C}erenkov muon
decay background shown in Fig.~\ref{fig:temp-spect} by a factor $\sim
5$, and would remove the spallation backgrounds in the range 10--18
MeV; below about 10 MeV, the reactor $\bar{\nu}_e$ signal suddenly
becomes dominant~\cite{GADZOOKS,Vagins}.
We emphasize that SK would be working well within its design range
when detecting these positrons and neutron captures, i.e., at these
energies the detection efficiency is very high, nearly constant, and
well-measured through calibrations.  In contrast, in the detection of neutrinos
from SN 1987A, both Kam-II and especially IMB had events detected at
energies where the efficiency was low and/or varying quickly~\cite{1987ADETECT}.

Besides increasing the signal rate
and improving the ability to test the DSNB spectral shape, the capability
for neutron tagging could allow a rate-limited, instead of background-limited,
search, so that the sensitivity could improve linearly with detector exposure.
To examine the prospects for a gadolinium-enhanced SK, we again consider
the four points in Table~\ref{table:temp-table}.  For each, we
simulate the expected neutrino spectra over a 5-year period (i.e., 5
times the yields shown in Fig.~\ref{fig:temp-spect}, noting the
background reduction).  We fit the spectra of DSNB and background
events simultaneously, and compute the $\chi^2$ in 4-MeV bins, as in
Ref.~\cite{Malek}.   With these statistics, the Gaussian $\chi^2$ is
adequate to draw the 90\% C.L. contours.

In Fig.~\ref{fig:temp-expect}, we show the expected determinations of
the physical parameters at 90\% C.L.  The spectra corresponding to
\textbf{A}, \textbf{B}, and \textbf{C}, while presently
indistinguishable above 18 MeV, would be clearly separable in a
gadolinium-enhanced SK (see Fig.~\ref{fig:temp-spect}).  With a lower
energy threshold, there is greater sensitivity to both the time-integrated
luminosity and average energy through the spectral shape as well as
normalization.  At different points, the relative sensitivity to the
two parameters changes, as listed in Table~\ref{table:temp-table} and
shown in Figs.~\ref{fig:temp-spect} and \ref{fig:temp-expect}.  If we
allowed the supernova rate history to be free, then the normalization
of the supernova rate would be degenerate with the time-integrated
luminosity, while the rate of increase with redshift would be
degenerate with the average energy.  Thus near point \textbf{A},
inaccuracies in the normalization would have little impact, and
likewise near point \textbf{C} for inaccuracies in the rate of
increase with redshift.  While we think that the supernova rate
uncertainties will play a minor role compared to those on
the supernova emission parameters, this may make the determination of
one variable more robust than the other.

If the true parameters, even taking neutrino mixing into account, are
closer to the canonical values (point \textbf{D}), then the detection
rate will be lower, and hence the allowed region larger.  (From the
size and shape of this contour relative to the others, one can estimate
how the contours for other points would look.)  Even in the case of point
\textbf{D}, in 5 years a gadolinium-enhanced SK would detect $\simeq 15$
events, comparable to the SN 1987A yield, which should help
distinguish between the Kam-II and IMB solutions.  If the Kam-II
region is correct, then the detection rate will be even lower; note
that an outcome that disfavored the IMB region might be viewed as
selecting the Kam-II region, given the prior information from SN 1987A.
If the supernova neutrino emission parameters are both low, a
megaton-scale detector may be necessary to accumulate good statistics.

%%%%%%%%%%%%%%%%%%%%%%%%%%%%%%%%%%%%%%%%%%

{\bf Discussion.---}%
The $\simeq 20$ neutrinos from SN 1987A provided a rough confirmation
of core-collapse supernova neutrino emission, and hence of the
dynamics of the exploding star in the first several seconds after
collapse.  These details of core-collapse events will forever remain
invisible with photons, but can be revealed by neutrinos, if they can
be detected.  While very challenging, it is hard to overstate the
importance of this goal.  Since nearly 20 years after SN 1987A, we
have no further {\it direct} information on supernova neutrino
emission, techniques besides waiting for a Milky Way supernova must be
considered.  

We propose that with the rapidly improving precision of
the astronomical data, it will be possible to use measurements of the
DSNB to constrain the $\bar{\nu}_e$ emission per supernova.  This
information will be limited, in that neither the dynamic timescales
nor the emission in the other neutrino flavors can be measured;
only a Milky Way supernova can provide those data.
However, even with just $\bar{\nu}_e$, the time-integrated luminosity and
average energy will constrain the explosion energy and proto-neutron
star opacity, especially if reasonable assumptions are made about the
other flavors.  Despite these caveats, and the limited statistics, we
stress that this technique is unique in that in a very short time it
could begin providing steadily better clues to the mysteries of SN
1987A.

Our results in Fig.~\ref{fig:temp-expect} show that a
gadolinium-enhanced SK detector would have useful sensitivity to
an interesting range of supernova emission parameters.  The recognition
that the astrophysical uncertainties are small and quickly diminishing
allowed us to reinterpret the SK flux constraint~\cite{Malek} in terms
of the neutrino emission per supernova.  As a practical matter, we
encourage SK to represent future results in this way, as it is more
directly connected to the measured event spectrum than the integral
of the flux above a given energy (in particular, since the latter does not
contain the important weighting by the detection cross section), i.e.,
is less model-dependent.

If a Milky Way supernova is detected, this would {\it increase} the value
of the proposed DSNB measurement.  The comparison
of results could probe whether the neutrino emission from
core-collapse supernovae is as uniform as presently assumed.
Alternatively, it could test the measured core-collapse rate history
\cite{Ando2004}, and whether there is an additional neutrino
background from explosions which fail~\cite{Concordance}, emitting
neutrinos but not creating an optical supernova, just as is seen in
simulations~\cite{SNsim}.

Proposed megaton-scale detectors would greatly extend the sensitivity
to these and more general spectra, and could bring precision to the
measurement, due to the very high DSNB statistics.
Such detectors could also allow the accumulation of events from identified
supernovae within 10 Mpc~\cite{PEEPING}.
While the statistics of the latter would be relatively lower, 
that spectrum would not be redshifted, nor dependent on the evolution
of the cosmic supernova rate, and hence would be complementary to
the DSNB spectrum.

The neutrino emission per supernova is also important for
understanding nucleosynthesis, especially of the heavy elements beyond
iron, which are believed to be formed only in core-collapse
supernovae, and which require special conditions that may be
importantly affected by the neutrinos~\cite{rproc}.  In addition,
Yoshida {\it et al.}~\cite{Yoshida} have recently shown that the yield
of the light element $^{11}$B constrains the neutrino emission
parameters to be close to the canonical values, which is favorable for
confirmation by {\it direct} detection.  Combining the nucleosynthesis
results~\cite{Yoshida} with future sensitivity to the DSNB electron
antineutrino flux (as stressed here)~\cite{Vagins,AST,DSNB,Ando2004,Concordance},
the DSNB electron neutrino flux~\cite{DSNBnue1,DSNBnue2}, and the summed
spectrum of nearby supernovae~\cite{PEEPING} will provide complementary
and restrictive probes of the details of supernova neutrino emission and the
history of stellar birth, life, and death.

%%%%%%%%%%%%%%%%%%%%%%%%%%%%%%%%%%%%%%%%%%

\vspace{6cm}

{\bf Acknowledgments.---}%
We thank L.~Strigari and M.~Vagins for discussions.  This work was
supported by The Ohio State University and NSF CAREER grant
No. PHY-0547102 to J.F.B.; S.A. was also supported by the Japan
Society for the Promotion of Science.

%%%%%%%%%%%%%%%%%%%%%%%%%%%%%%%%%%%%%%%%%%
%%%%%%%%%%%%%%%%%%%%%%%%%%%%%%%%%%%%%%%%%%

%%%%%%%%%%%%%%%%%%%%%%%%%%%%%%%%%%%%%%%%%%
%%%%%%%%%%%%%%%%%%%%%%%%%%%%%%%%%%%%%%%%%%

\end{document}